\documentclass[a4paper]{spie}  
\usepackage[]{graphicx}
\newcommand{\degr}{$^{\circ}\,$}

\title{GBOT - Ground Based Optical Tracking of the Gaia satellite}

\author{Martin Altmann\supit{a,b}, Sebastien Bouquillon\supit{b}, 
Francois Taris\supit{b}, Iain A. Steele\supit{c}, 
Ricky L. Smart\supit{d}, Alexandre H. Andrei\supit{b,d,e,f}, 
Christophe Barache\supit{b}, Teddy Carlucci\supit{b} 
and Sebastian G. Els\supit{a,g}
\skiplinehalf
\supit{a}Zentrum f\"ur Astronomie d. Univ. Heidelberg, M\"onchhofstr. 12-14, 69120 Heidelberg, Germany\\
\supit{b}SYRTE, Observatoire de Paris, Av. Denfert-Rochereau 77, 75014 Paris, France\\
\supit{c}Astrophysical Research Unit of the John Moores University, 146 Brownlow Hill, Liverpool L3 5RF, United Kingdom\\ 
\supit{d}INAF Osservatorio Astrofisico di Torino , Pino Torinese, Italy\\
\supit{e}Observat\'orio Nacional, Rua General Jos\'e Cristino 77, S\~ao Cristovao, Rio de Janeiro, RJ 20921-400, Brasil \\
\supit{f}Observatorio de Valongo, UFRJ , Rio de Janeiro, Brazil\\
\supit{g}European Space Astronomy Centre (ESAC), Camino Bajo del Castillo, s/n., Urb. Villafranca del Castillo, 28692 Villanueva de la Ca\~nada, Madrid, Spain
}

\authorinfo{Further author information: (Send correspondence to Martin Altmann)\\Martin Altmann: E-mail: maltmann@ari.uni-heidelberg.de, Telephone: +49 6224 541818}

\begin{document} 
  \maketitle 

\begin{abstract}
Gaia, the 1 billion star, high precision, astrometric satellite will revolutionise our 
understanding in many areas of astronomy ranging from bodies in our Solar System to the formation 
and structure of our Galaxy. To fully achieve the ambitious goals of the mission, 
and to completely eliminate effects such as aberration, we must know the position 
and velocity vectors of the spacecraft as it orbits the Lagrange point to an accuracy greater 
than can be obtained by traditional radar techniques, leading to the decision to conduct astrometric observations 
of the Gaia satellite itself from the ground. Therefore the 
Ground Based Optical Tracking (GBOT) project was formed and a small worldwide network using 1-2 m 
telescopes established in order to obtain one measurement per day of a 
precision/accuracy of 20 mas. We will discuss all aspects of GBOT, setup, feasibility considerations, 
preliminary tests of observing methods, partner observatories, the pipeline/database (see also contribution by Bouquillon et al.\cite{Bouquillon_SPIE2014}).

\end{abstract}



\section{INTRODUCTION}
\label{sec:intro}  
ESA's Gaia astrometric satellite mission, launched on December 19, 2013 will revolutionise our view of 
the structure, formation, and evolution of our Milky Way, and therefore of galaxies in general. Additionally it will make significant
contributions to several other fields of astrophysics and physics, from the foundations of basic physics, small solar system objects, exoplanets 
to cosmology. It will provide high quality absolute astrometry, i.e. proper motions, positions and parallaxes for 1 billion stars - 
every object brighter than 20th magnitude! Furthermore using two low resolution spectrophotometric devices, it will obtain multipassband photometry
for the complete sample, and using a high resolution spectrograph, radial velocities and astrophysical parameters for a subset of brighter stars. 
For a more complete presentation of the capabilities, goals, and current status of the Gaia mission see the plenary presentation by Timo Prusti\cite{Prusti_SPIE2014}.

Given the unprecedented precision\footnote{Throughout this article, precision and accuracy are strictly kept separate, unlike in many cases, where the two terms are often interchanged. Precision
refers to scatter, accuracy to offsets} of Gaia's measurements and the fact that it is a global mission, i.e. not focussing on small fields, but the complete
sky, there are a number of effects that need to be taken care of, such as gravitational deflection, and aberration. 
In some cases the conventional means no longer suffice, and new approaches need to be considered. In the precision level
of Gaia two problematic effects are:
\begin{itemize}
\item {\bf Aberration:} This effect is in the order of tens of arc seconds. It is corrected for by determination of the motion vector of the satellite
(or the telescope on Earth for ground based global astrometry) in respect to a well defined reference point, usually the barycentre of the Solar 
system. In the case of the preceding Hipparchos mission\cite{Hipparchos},
 the conventional methods of determination using the ranging and
communications stations data were sufficient, but with Gaia we enter 
a realm, in which this is no longer always the case. While for most stars and most of the time the aberration can still be corrected by inserting the
ranging station into the orbit reconstruction, this is not possible
 for those bright stars (12-15 mag) which have the most precise measurements
and in some parts of the orbit, especially when the space craft crosses the celestial equator and after major corrective boosts. 

Another reason of minimising
effects such as aberration even in those cases, for which the residua are not detectable for one object. The vast number of stars observed by Gaia will allow the composition
of large samples of even relatively rare stellar types which in turn allow a very precise determination of properties like e.g. typical luminosities. One would not want to 
compromise the accuracy of such highly precise results unnecessarily by systematic effects, such as aberration.     
\item {\bf The baseline for parallax measurements of nearby solar system objects:} 
The geometric situation when measuring a solar system object is a very different one that when dealing with an extrasolar source. Since
the target is moving fast, the distance is comparable with the baseline, 
and the baseline is much smaller than in the classical stellar case (for which the baseline is in the vicinity
of the Sun -Earth distance), the distance to the target is in the order of $10^6-10^9\times$ the baseline. For a solar system object, the measurements are usually made about 2 hours apart
(with the second FOV passing the field), during which the object itself has moved significantly. 
The baseline is accordingly only in the order of 100,000 km and the distance of the 
target 100-1000$\times$ the baseline. This means that we need to know the length of the baseline very precisely, i.e. we need to be able to determine the position of Gaia to a very high 
degrees of precision.
\end{itemize}
These two effects lead to the specification that the velocity of Gaia must be known to better than {\bf 2.5 mm/s} and the position to better than {\bf 150 m}. This translates to 
a positional precision of 21 mas every day, and a motion of 29 mas/d. To ensure this 
it was decided that new approaches to this problem will be implemented, since one single ranging station cannot guarantee this at all times, and two ranging stations are usually not an option,
given their scarcity and high price. This lead to proposal to use ground based telescopes of moderate size to conduct an astrometric campaign on the Gaia  satellite itself and
the conception of the GBOT (Ground Based Optical Tracking) project. While Ground Based Satellite tracking projects are not new, in fact they have been around since Sputnik in 1957, 
see e.g. Masevich \& Losinskii\cite{Russat}, the novelty is
a high precision and accuracy astrometric tracking campaign ( at a level of 20 mas on a daily basis) of a spacecraft, 
i.e. a very faint (given its location in the Sun Earth L2 point region) and moving object on a typically daily cadence for a timespan of 5 years or more (the duration of Gaia's operating phase). 
This article reports on the setup and structure of the GBOT project, the theoretical considerations to take into account, the requirements for participating telescopes and observatories, 
testing and the initialisation of the actual operational phase, which in the case of Gaia has lead to the need to reassess the whole campaign due to the unexpected faintness of the target.
The software pipeline explicitly developed for this project and the underlying database system, will only briefly be touched in this paper, a more comprehensive and detailed version is given
in another article dedicated to those components (Bouquillon et al. 2014)\cite{Bouquillon_SPIE2014}.

The experience we have gathered with developing, installing and operating GBOT, could possibly serve as a template or recipe for future astrometric projects, or other projects requiring a similar degrees
of of knowledge of the state of the 6D vector of the spacecraft in question. Assuming that a future global astrometric space mission, should this ever be attempted will undoubtedly aim for
even higher precision data than Gaia does today, a GBOT-like campaign may even move from ``important'' to ``essential''. Similar techniques are also being used by ``space debris'' tracking programs, 
and indeed, several groups have stated interest in our project. 

GBOT is part of Gaia's {\it Data Processing and Analysis Consortium}, a body of about 400 scientists charged with the operations of the Gaia mission and the reduction of the data. 

\section{THE GBOT CONCEPT}
\label{sec:concept}

\begin{figure}
   \begin{center}
   \includegraphics[height=10cm]{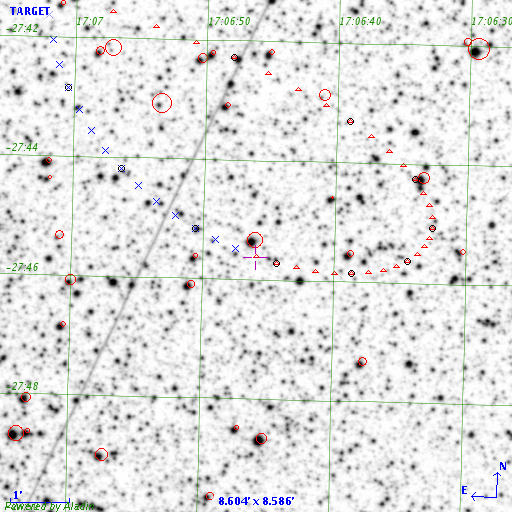}
   \end{center}
   \caption[Gaia - typical trajectory in the sky]{
   \label{fig:gaiasky}
This figure shows the track of Gaia in the sky (in this case for Paranal Observatory, May 26, 2014) The crosses and triangles denote the track of Gaia before and after the time of closest zenithal 
distance respectively. The circled stars are 2MASS\cite{2MASS} stars used for the field identification. This chart was made using the GBOT FOV tool, see 
Bouquillon et al. (2014)\cite{Bouquillon_SPIE2014}, see their Section 3 and Figure 2
}
   \end{figure}

Following the requirements of 150 m in position and 2.5 mm/s in velocity, a set of requirements for the GBOT project was set up. Since earth bound astrometry can only 
produce information about the transversal motion and 2D positions in the sky, the third component, i.e. the radial velocity and distance must come from somewhere else. Fortunately, the
classical radar ranging easily fulfils all requirements in the radial direction, i.e. for Gaia there is no need to discuss better alternatives for this dimension\footnote{This may be another matter 
in the case of tracking campaigns for other 
missions with other requirements in precision in the future, however this discussion is beyond the scope of this article.}. 

A long term tracking program such as GBOT, which is projected to last during the whole operational phase of Gaia (5-6 years)
requires a set of basic
principles and requirements in order to function properly. Based on qualified assumptions, an estimate of the putative brightness of the target was made. This is one of the most important
constraints, and at the same time one of the most uncertain, as the GBOT group had to learn, see Sect. \ref{sec:operations}. The target brightness determines the 
range of suitable telescope facilities. In this case (see Sect. \ref{sec:theory}), an average brightness of $R\simeq$18 mag was estimated. The GBOT program was conceived at a phase where the design
of the Gaia spacecraft itself was principally complete, for future missions, requiring ground based astrometric tracking,
 it will be very worthwhile to consider the needs of such a program, by studying the
reflective properties more closely and if possible ensuring a viable brightness in the order of about 17-18 mag. This is 
the ideal magnitude range as a compromise between object S/N and
number of background reference stars not too different in brightness. Measures such as a light emitting source (LED array) or a cat's eye reflector might be considered as possible brightness enhancers.

Assuming a brightness of 18 mag in red passbands, the requirements of GBOT on partner observatories were:
\begin{itemize}
\item {\bf Telescope diameter:} at least 1 m. Therefore GBOT primarily looked for facilities in the 1-2 m range. The GBOT reassessment necessary due to the far fainter brightness of 21 mag after launch,
came to the conclusion that 2 m telescopes are still able to deliver the necessary data, possibly by increasing the number of exposures per sequence.
\item {\bf Detector properties:} The available detector needs to be able to provide adequately sampled data, i.e. the pixel scale must be above a certain limit. Since in most cases the S/N is
the key limit to precision a certain level of undersampling can be tolerated; GBOT agreed on a limit of 0.4$\times$median seeing. The Field of view also plays an important role, since the
astrometric reduction requires a certain number of reference stars, and the more parameters to be accounted for (geometric term of first, second, third or even higher order, magnitude dependent terms,
colour terms, such as differential colour refraction (DCR), etc.) the more reference stars are required. The minimum FOV for GBOT observations was chosen to be 5'$\times$5', with 10'-15'$\times$10'-15'
 being optimal. Larger FOV's are also acceptable, but they often have the disadvantage, that they are either realised using mosaic detector arrays, or have a too coarse pixel scale. Additionally
effects such as differential refraction (DR) or even differential aberration start playing a role with large fields. Given the inherent problems of obtaining high accuracy astrometry from
CCD mosaics, GBOT only uses monolithic data: While GBOT does use data from such arrays, all but the frame containing the target are discarded. 
\item {\bf Facility infrastructure:} The facility needs to be able to provide data on a regular basis. Therefore the prime option are robotic telescopes working off a queue of observations autonomously. Also very suited
are telescopes used for long term monitoring campaigns, or mainly used in queue observing mode. Less suited are telescopes mainly used by visitors, since this means negotiating with each new
visitor. This requirement does not only encompass the frequency of data, but also the availability of new data for download and the injection into GBOT's system. Therefore high band Internet and preferably
and archival system is necessary. GBOT prefers data which has been detrended, i.e. the basic image reduction steps have been carried out. However this is not a major concern, as routines performing
the detrending can easily be implemented. 
\item {\bf Metadata constraints:} Finally there are a number of constraints on the metadata that accompanies the actual data, namely clock accuracy and the telescope's geographical coordinates. The former must be better
than 0.1 secs absolute, and the latter better than 7 m per direction\footnote{For other space missions these might be different, but note that the time constraint cannot be significantly better than
this, even if the clocking system implies this, since the effective time is also influenced by the time-distribution of the light falling onto the detector, and this is influenced by thin clouds etc.}.
The geographic coordinates can nowadays easily be determined using GPS measurements done repeatedly; for the altitude precise maps may also be of help, since GPS is less precise in this coordinate. The 
timestamp accuracy issue is more complicated since this does not only depend on the clock accuracy, but also the time delay a command documented by the timestamp in the image file's header gets
passed on to the shutter. Moreover a shutter also needs a finite time to open and close, so close attention has to be given
 to the characteristics and types of shutter and whether they open/close
symmetrically or asymmetrically. A higher accuracy than 0.1 sec would be desirable, but is not realistic since, even if the technological ability of the telescope/instrument facility setup
would support this, minor transparency variations could shift the effective time of an exposure. 
Therefore we decided to limit our requirement in this case on the minimum possible. For 
future projects with possibly higher requirements in this regard, the influence of ambient conditions on the effective observing time need to be investigated.  
\end{itemize}

In order to evaluate a prospective facility, a set of standard tests was developed, and the suitability documented. These include test observations of 
an L2-spacecraft (before Gaia this role was played by WMAP and Planck), under different ambient conditions (seeing, transparency, etc.) 
and lunar phases, and astrometric tests, with a specific star field being observed
and reobserved with the pointing changed by half a field of view. For those partners selected, we set up an ``Interface Control Document'' (ICD), which clearly and concisely 
documents all particulars (such as observing sequence details, and cadence) and agreements and communication and data download 
procedures concerning the data acquisition process done with this facility. 

Finally, since 2012 we are also developing another observing campaign using VLBI radio astrometry techniques, dubbed ``Radio-GBOT''.
 This has the advantage, of much more precise measurements with the drawback that
the realistic cadence is much lower. We hope that both approaches complement each other, with each playing out its strengths. However a full description of this effort is beyond the scope of this
paper which focuses on the optical campaign.

\section{THEORETICAL CONSIDERATIONS}
\label{sec:theory}
Before starting to organise and set up a GBOT-like campaign a few theoretical basics need to be considered.
There needs to be certainty that the constraints set by the mission in terms of precision and accuracy can be met.
This does not only apply to the data itself and its quality but also other parameters, such as time stamp accuracy and the
values for the geographic coordinates (see Sect. \ref{sec:concept}).

In the case of GBOT, we face a problem which similar post-Gaia efforts will most likely not encounter: the absolute
proof that the required quality can be reached can only be given, 
once the GBOT data has been reduced with Gaia astrometry as reference material. For the future
Gaia data will be available from the start. Current material, such as the PPMXL\cite{ppmxl}, etc. have a significant zonal error problem, caused by the way they are assembled.
Moreover the precision at 18+ magnitudes (the range of most reference stars in a GBOT field) is about 200-300 mas (compared with 0.5 mas in the case of Gaia). This and the systematics, which can 
amount to up to 100 mas, will lead to systematic fit errors of the magnitude 50-100 mas. 
Therefore when comparing our results with a reconstructed orbit (i.e. Observed-Calculated
analysis) the mean O-C is not very much representative of anything, however the scatter does give a relatively good indication about the precision although there are other effects than
true r.m.s. scatter visible in many cases. 

One of the more important parameters to consider is the object tangential velocity, since this effect limits the maximum exposure time per exposure. 
Gaia is located in the L2, i.e. ideally it would take the same time to orbit the Sun as Earth, giving it
a  mean velocity of 1\degr per day. This is not entirely true, since Gaia is on a Lissajous type orbit taking it up to 15\degr away from the nominal L2 point, the distance covered is larger 
and so is the average velocity. But the most important effect is Earth's rotation, and fortunately this mostly plays in the favour of the observer. At midnight Earth ``overtakes''
the L2 and thus the target so that the tangential velocity drops significantly, and can even go down to values close to zero very near to the spacecraft's culmination. The average 1\degr/day 
velocity translates to 42 mas/sec, and during the observing period in each night, i.e. an hour angle of less than 2.5 hrs. it is generally less than 30 mas/sec, a much more manageable amount for 
tracking campaigns. A small downside of this effect is the fact that especially during the night, the object does not move in a straight line, but describes as loop on the sky plane, as shown in 
Fig. \ref{fig:gaiasky}.
This means that GBOT delivers every single measurement to SOC, instead of only an average of all of a sequence. However, 
the precision 
commitments made by GBOT for 20 mas per day, applies to the whole 
sequence rather than to an individual exposure. In general the maximum useful exposure time for GBOT is in the order of 60 secs, which results only in a slight elongation of about 1''.   
\subsection{The Gaia brightness issue}
\label{sec:theory:brightness}
  \begin{figure}
   \begin{center}
   \includegraphics[height=8cm]{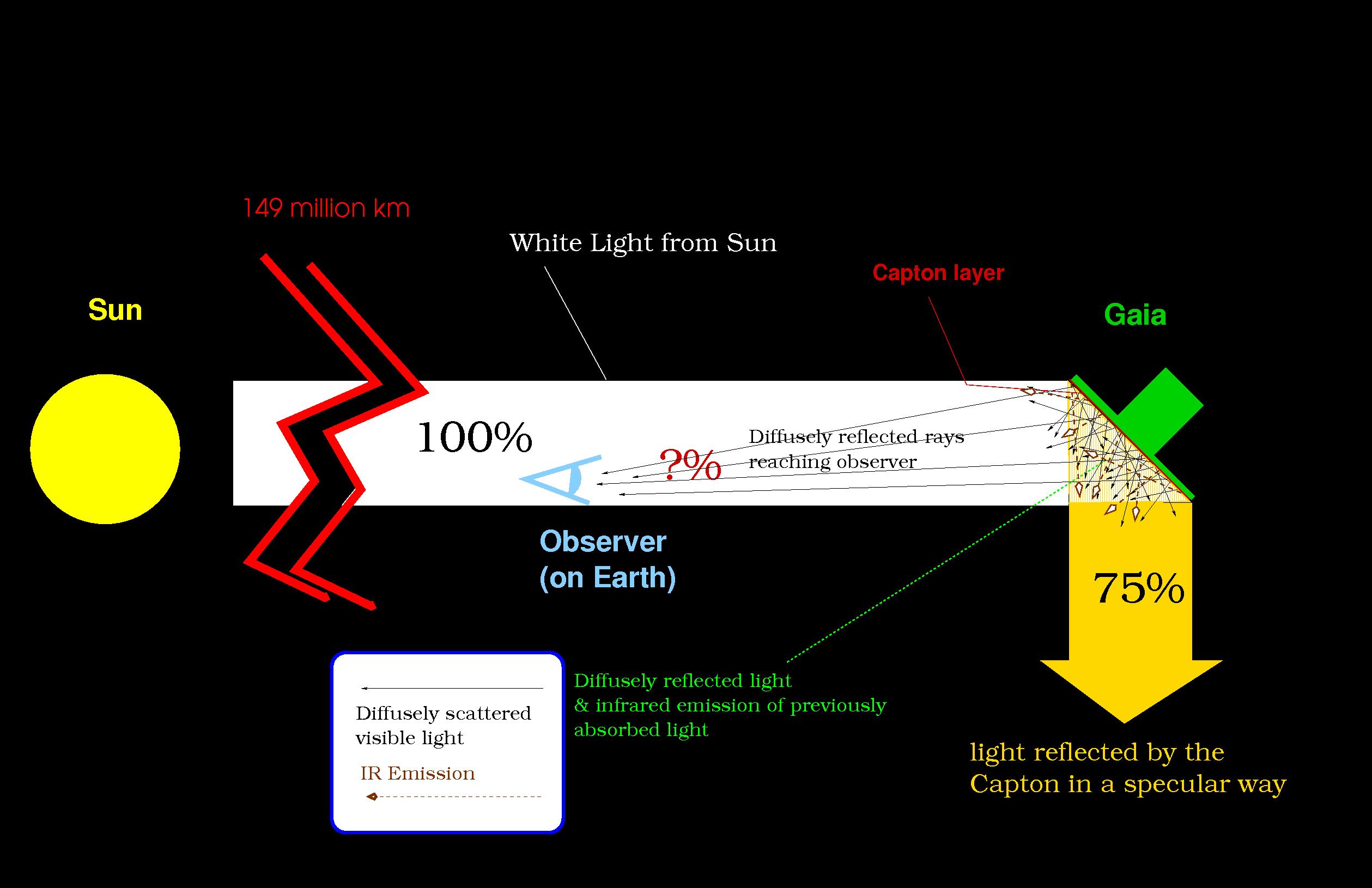}
   \end{center}
   \caption[Gaia - light path geometry]{
   \label{fig:gaiarefl}
  The path of light from the Sun via Gaia to the observer on Earth. As can clearly be seen, we have to rely on diffuse scattering, caused by wrinkles, cracks, impurities, of the Kapton layer, and
other structure, such as communication and data transfer antennae, etc.}
   \end{figure}
  The brightness of the main target of a campaign such as Gaia is probably the single most important piece of information. Unfortunately it is also one of the hardest to obtain, since in most
cases the satellite is being assembles on the ground while the campaign is set up. While it looks rather straightforward at first to draw some conclusion based on basic principles, such as 
area, reflectivity, etc. in practice this is a very complicated process, since it depends on the reflective quantities of the material, which are not always available because of commercial considerations, and
may furthermore change over time, due to effects like radiation damage - the L2 is a high radiation environment. 
Other factors are the aspect angle, the azimuth angle (which is constantly changing for a
spinning object like Gaia), and many other issues.  
We had to
 rely on our experience with other space craft, namely 
those we were observing during our tests, i.e. mainly WMAP and 
Planck; While Planck has a widely different shape and probably also reflective characteristics, 
see Fig. \ref{fig:grond}, WMAP was essentially a smaller version of Gaia inclined by half the angle
of Gaia, i.e. 22.5\degr instead of 45\degr. WMAP in general had a magnitude in $R$ of 18-18.5, 
Planck about 18 mag. Therefore it seemed safe to assume that, Gaia would have 
roughly the same magnitude. This turned out to be not the case. For reasons not yet understood,
 Gaia turned out to be more than 2 mags fainter than Planck, leading to
the need to reevaluate the whole GBOT program, as described in Sect. \ref{sec:operations}. 
\subsection{Precision}
\label{sec:theory:precision}
   \begin{figure}
   \begin{center}
\vspace*{-1cm}
   \includegraphics[height=18cm]{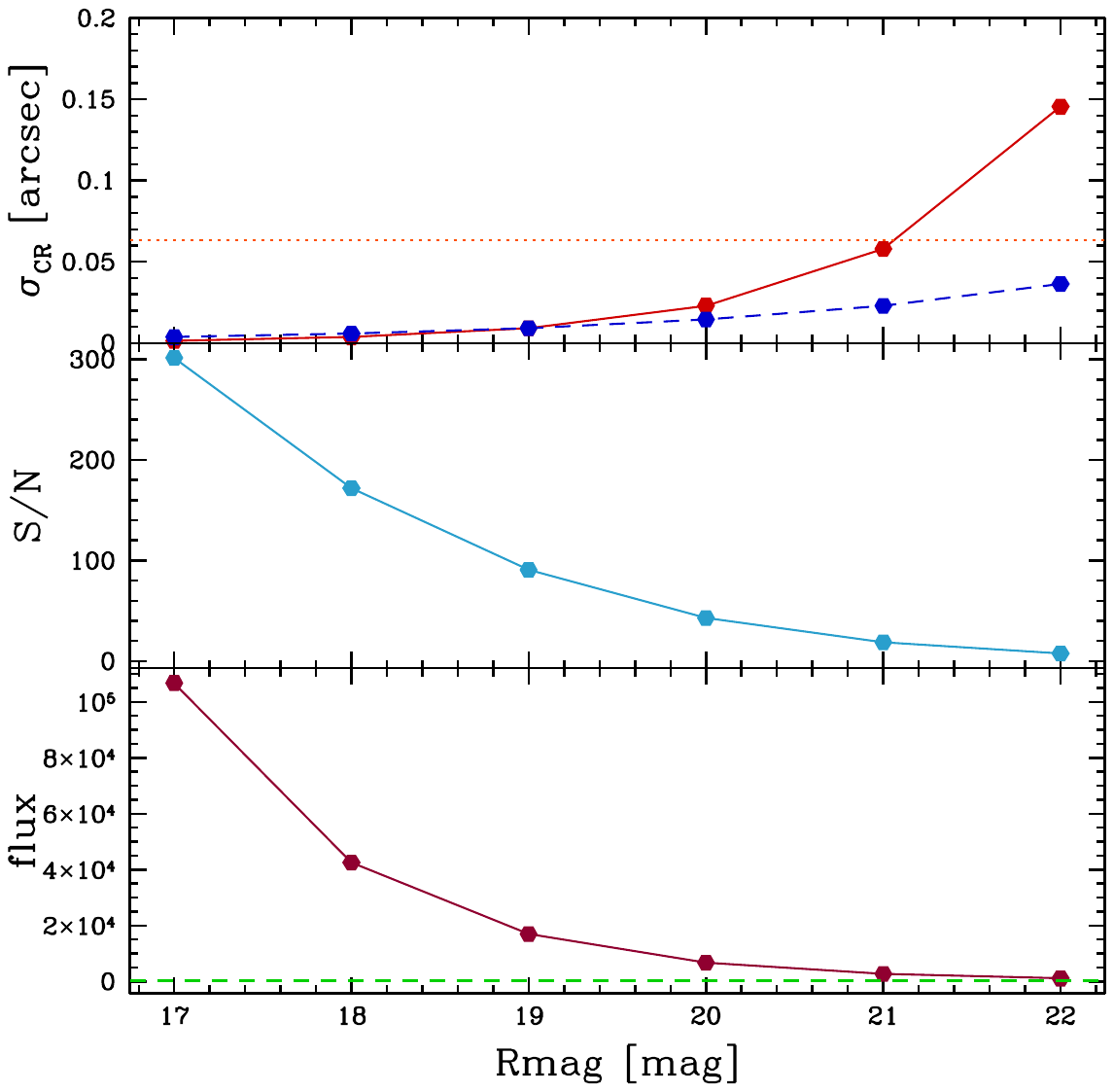}
   \end{center}
\vspace*{-3cm}
   \caption[CRLB in relation to magnitude for the 2.2 m ESO telescope]{
   \label{fig:CRLB}
Cramer Rao limits for stars of various magnitudes as observed with the ESO-La Silla 2.2 m and its WFI detector. The count rates/fluxes were 
derived from the instrument's Exposure time Calculator on the ESO website. The input parameters kept constant are: exposure time 60 secs, $R_{\rm SDSS}$ filter,
seeing 1.2'', Airmass 1.6'',
lunar phase 3 days (resulting in a sky background of 254.87 $e^-$). The lower two panels show the S/N, and total flux, as taken from the 
ETC, the green  dashed line indicates the background flux. The upper panel shows the Cramer Rao limits calculated using the approximations of
Mend\'ez et al. (2013)\cite{Mendez1} (Their equations 39, 42, 45). The blue dashed curve depicts the high S/N ($Flux >> BG $ approximation, the red solid curve the low 
(i.e. $Flux < BG$) approach in dependency of the total flux. In the middle part, neither approximation is strictly valid, however, they are very similar in this region, therefore as a 
conservative approach, one should take the maximum of the two curves. The orange dotted (light grey in printed version) line denotes the value for $\sigma_{\rm CR}$ which produces a r.m.s. error
of 20 mas if a sequence of 10 exposures is used}
   \end{figure}

The source extraction precision defines the minimum of error reachable, i.e. the maximum quality. 
This aspect becomes extremely important if one observes in the very faint regime especially when the object is fainter than the background. The main
parameter for the precision is the amount of object- and unwanted flux gathered on the chip, i.e. the S/N. This is constrained by the brightness of object and sky,
the CCD-read out noise, and the exposure time, the latter being limited by the object motion. Assessment of the achievable precision is all the more important the fainter the target object is, especially
once the target brightness gets close to the background brightness or even drops below it. 
In the case of GBOT, the pre-launch scenario of a magnitude of $\simeq18$ mag meant that Gaia, like the test
objects Planck and WMAP, would be 2-3 mags brighter than the usual sky background at a typical observatory. The actual magnitude, as found out post-launch is $\simeq20-21.2$ mag, i.e. comparable to the
ambient sky brightness. This lead us to study the limits of precision much more closely during the GBOT reassessment process fully described in Sect. \ref{sec:operations}.

The lower limits of the variance of an estimate in terms of its estimators is defined by the Cramer-Rao lower bound (CRLB, Cramer (1946)\cite{Cramer} and Rao (1945)\cite{Rao}). Parameters flowing into the estimate for astrometric precision, in
our case are signal strength, i.e. S/N, background level, pixel scale, FWHM. In a recent publication by Mend'ez et al. (2013)\cite{Mendez1} have intensively studied the CRLB for 1D-CCD data, our analysis strongly depends
on that work. Fig. \ref{fig:CRLB} shows the situation for a 2 m telescope in a good location, in this case the 2.2 m MPIA telescope located on ESO's La Silla observatory, and its WFI mosaic detector.
S/N and background values have been computed using the ESO WFI exposure time calculator and the according Cramer Rao limits were computed using equations 39, 42 and 45 of Mend\'ez et al. (2013)\cite{Mendez1}. The two lowest 
panels show two different representations of the flux, the second one from the top the S/N in relation to the magnitude and the top panel the CRLB. In this panel the upper pair of curves (red in the online version) denote the faint object approximation, i.e. the situation relevant to GBOT, and the lower pair (blue in the online version) the bright object approximation. The dotted line in the panel 
near 63 mas shows the scatter which will lead to a precision of 20 mas for a series of 10 exposures of 60 secs, i.e. the nominal GBOT sequence. For this telescope the CRLB is just sufficient
for the aim of 20 mas; therefore extending the nominal sequence to 12 or 15 exposures is recommended. For the ESO 2.6 m VST located on Mt. Paranal in Chile the situation is better, indeed in most cases
10 exposures will suffice. At this point it needs to be stressed, that the CRLB reflects the theoretical optimum of what can be achieved, actual centroiding methods can approach the CRLB but not
reach it. In practise, specially in the faint Gaia scenario, we are optimising our centroiding methods to approach the CRLB.  

\subsection{Accuracy issues and systematic effects}
\label{sec:theory:accuracy}

Apart from the S/N-driven precision, the results are influenced by a number of systematic effects, which can negatively influence the astrometry, and need to be quantified and corrected as completely
as possible. In most cases the correction can be achieved by adding terms to the polynomial defining the astrometric field model - if the field contains enough background stars. For small fields, this
may be a problem, especially in sparse fields at high Galactic latitude. Fortunately for most effects the systematic detrimental influence correlates with field size. 
The main systematic effects are:
\begin{itemize}
\item {\bf Optics:} Correcting for optical effects, such as distortion, chromatic aberration, astigmatism, defocus, and their effects on the optical field as pictured by an optical system are the core of what
is called astrometric reduction, and need to be corrected either by fitting general polynomials to the derived $x,y$ coordinates or by correcting the various effects explicitly by using the underlying
optics laws.
\item {\bf Atmospheric refraction:} This is a complicated effect, consisting of a colour independent term and a colour term. The absolute refraction can be a $>30'$ effect; 
however the absolute term is of no relevance at least in first order, since we do not have whole sky fields, so that the pointing of the 
telescope accommodates the absolute refraction. This is a little different in the second order effect, the differential refraction (DR), i.e. the difference in refraction in dependence of the zenithal distance. 
In principle this can indeed cause a systematic effect, in that stars in the lower part of the field are affected more by refraction than those on the upper part. 
Since in most cases right ascension and declination are not aligned to the azimuth and elevation, there are DR effects in both celestial coordinates. The DR is dependent on the zenithal
distance and the size of the FOV. Our theoretical considerations show, that for a 30' FOV, a linear correction is sufficient in most cases, to a zenithal distance of 50\degr, 
in the case of smaller fields, even lower. In most circumstances, the DR plays a rather minor role. This is different for the third part of refraction, differential colour refraction (DCR). This
arises from the very principle of refraction, namely that light of different wavelength is refracted by a different amount; thus the bulk of the light of stars of different colour is refracted 
by a different amount, leading to systematic differences of the positions of stars of different colour, hence a colour term. This must be corrected, and if possible minimised starting at the observing
process, by using redder filters, which are less affected by DCR. Unfortunately the reddest passbands at our disposal, namely $I/i$ type and $z$ have a couple of significant drawbacks, namely
the deteriorating QE-efficiency of most detectors in this wavelength regime, the high sky background, very much limiting the achievable S/N and fringing. Therefore we concentrated on $R,r$ type passbands, which show
a significantly larger amount of DCR, but also a better sensitivity. While there are several approaches of solving the DCR, often this effect can not be eradicated entirely. The usual method is to 
derive a colour term, i.e. a relation DCR-related offset vs. object colour. This implies that the colour of the target object and all reference objects are known, either by making exposures
in multiple passband during the actual observations or by using the colour information in the reference catalogues and a priori measurements of the objects colours. Using a second passband is usually
not possible on a daily basis, given the demand on the telescope time, and at current for most parts of the sky the colour 
information is based on photographic plates, and thus is not really reliable. Again this is one of the things which will change once the data from Gaia all sky photometry is available. Finally 
the DCR is also depending on the atmospheric conditions, such as temperature and air pressure, not only at the site of detection but across the whole light vector, something usually not easy to
accommodate in the solution.     
\item {\bf Flatfield effects and detector issues:} We have also analysed the impact of residual flatfield effects, i.e. detector sensitivity variations
 and detector defects on our astrometry. Objects on known bad columns or other 
heavily blemished areas on a detector should be rejected. This means that care is to be taken, that the target object is not located on any of these regions or even close to them. Likewise 
affected reference stars should be removed from the list of reference stars. The effect of small scale and large scale flatfield effects can be solved analytically. It turns out that position is
quite robust against these effects, more so than photometry. Mid-scale effects are more difficult to solve, we assume that the impact is similar to that of large and small features. Nonetheless care should be
taken that the field is as smooth as possible, a good flat field correction is highly desirable. 
\item {\bf background stars:} A background star over which the target object moves alters the locus of the target. While for background objects significantly fainter than the target,
 the impact is rather small, this is no longer the case for such objects having a flux of more than a few \% of the target. Of course this presents itself as a much larger problem for the
21 mag Gaia than it did for the 18th mag object Gaia was assumed to be. Bright stars in Gaia's path can in principle be marked and communicated to an observatory so that they are informed 
and do not observe in the problematic time since the number of bright stars which can be identified a priori is rather small and the whole process is complicated, GBOT does at present not
use any pre-warning system. This may change in the future, e.g. once Gaia data are available. The vast number of potential interlopers are faint and uncharted, therefore the results of the 
astrometric reduction and the 2D images need to be carefully inspected and a dubious sequence removed from the list of observations. Since Gaia in most times moves fast enough to completely transverse
the problematic zone within a 10$\times$60 sec sequence, a background stars usually shows up as a signal in the $O-C$.  
\item {\bf Other effects:} Effects like differential aberration, moon light, sky transparency also play a role. Additionally some non-optical issues are of significance for a program like GBOT. These
include time stamp accuracy, and geographical coordinates. The maximum tolerances are given in the GBOT specifications, see Sect. \ref{sec:concept}.
\end{itemize}
The GBOT project is taking care to carefully assess the impact of each of the issues described in Sections \ref{sec:theory:precision} and \ref{sec:theory:accuracy} and find ways to compensate or at 
least minimise their influence on the astrometry. Several internal studies\footnote{which have been made public, and can be accessed under:\\ {\tt http://www.cosmos.esa.int/web/gaia/public-dpac-documents}} have been made for this purpose. At current,
 the overwhelming systematic effects on our results are the insufficient quality reference catalogue data (both astrometrically and photometrically) and the DCR. As described
 in the previous Section, some of the most important effects, can only be taken care of after the first release of Gaia data, which will take place about 2 years after start of Gaia 
operations, i.e. some time in 2015. According to our study\cite{MA009}, the total budget of systematic residuals
 can be pushed to below 5 mas. We hope that future similar enterprises will face fewer challenges, with Gaia data being available from the beginning. 
\section{OBSERVING TESTS}
\label{sec:observations}
While a theoretical evaluation of what can be expected, such as described in Sect. \ref{sec:theory} is of great value, the results need to be confirmed with real observations. The findings can be 
different to some degrees from the expectations based on theory, since some unknown or unregarded parameters can also play a role. For this reason observational tests are inevitable. Lacking our
prime target in the sky (with Gaia being safely on the ground), we had to use proxy objects to gather our data. These have to be of similar magnitude as the assumed Gaia magnitude, and share a similar
motion in the sky. Obvious prime choices are other space probes in the L2 region. The number of these is limited, since usually only space science missions occupy this region, the use of this region
is relatively recent, and the L2 region, belonging to one of the unstable Lagrange points cleans itself of objects, if their orbits are not maintained. 
Fortunately we could use WMAP, Planck and Herschel as guinea pigs. Secondary test objects were some main belt asteroids with a similar transversal velocity and brightness, i.e. near $R=18$ mag. 
These were used mainly in case
the L2 region was positioned inconveniently, such as near the summer solstice of a given hemisphere. Test observations were made under all possible sky conditions, seeing and lunar phases.
Additionally we observed astrometric fields with different pointings (so that two images overlap by 50\%), in order to study the field structure of an instrument. At first these tests were done to
establish our method in principle; most of these tests were done with the 2 m Liverpool telescope on La Palma, Canary Islands, supplemented by the 2.2 m MPIA telescope +WFI located on ESO's
La Silla telescope in Chile. When this was complete, a set of standard test observations containing those tests outlined above, were set up and given to every prospective contributor.   

Before the launch of Gaia, DPAC carried out a number of rehearsals, with GBOT taking part in two of these. The reason for these operations rehearsals (ORs) is to verify that the different components of
the Gaia operations are able to work together in a concerted way, something inevitable for such a complicated mission as Gaia. For GBOT, apart from the official aim, these were opportunities
to test the whole system for a prolonged time, i.e. 10-14 days. In both instances the test satellite was Planck and in OR2 GBOT worked only with one observatory, the LT, and in the subsequent
OR3, the LCOGT network joined in. The data accumulated during these test runs provided a valuable dataset for a detailed analysis, which for OR3 was documented in a technical note\cite{MA013}.
\section{GBOT: THE STRUCTURE OF THE  NETWORK}
\label{sec:setup}
\subsection{The Partner Observatories}
\label{sec:setup:tel}
The original setup of GBOT consisted of 1-2 m class telescopes, at the time of Gaia's launch, these were the 1 m telescopes of the Las Cumbres Optical Global Telescope Network (LCOGT),
which are located in Siding Spring, Australia (2 units), Sutherland (SAAO), South Africa (3 units), Cerro Tololo (CTIO), Chile (3 units), Mac Donald Observatory in Texas, USA (1, later 2 units),
the 2 m Liverpool telescope (LT) on Roque de los Muchachos, La Palma, Canary Islands, Spain, the VLT-Survey-Telescope (VST, ESO) at Mt. Paranal, Chile, as main contributors. Two further facilities,
the Euler 1.2 m telescope on La Silla, Chile, and the 1.06 m telescope at Pic du Midi, France were recruited as backups. These telescopes all fulfil the set of constraints outline in 
Sect. \ref{sec:concept}. Due to the faintness of Gaia, the setup, which is not finalised yet, looks
somewhat different. The VST and LT currently provide the backbone of GBOT's data acquisition, soon to be followed by LCOGT's 
two Faulkes 2 m telescopes, located at Siding Spring and Mauna Kea, Hawaii.
Some further facilities are being investigated. GBOT now relies on telescopes of the 2-3 m class. The robotic telescopes (LT, LCOGT) 
are operated using an observing queue, with the GBOT observation
being one of many observing programs. They will, ambient conditions permitting, observe one standard sequence per night, the LCOGT network acts as one telescope, i.e. it is internally scheduled
which unit carries out the observations. VST is also operated in queue mode, but the modus operandi is a little different. Here an ESO staff astronomer carries out the observations and manually
feeds an observing block (OB) into the queue, chosen according to ambient conditions, programme ranking, etc., i.e. a decision making process in principle similar to that of the robotic telescope but with 
human interaction. The main difference for the observer is that, while for the robotic telescope there is one observing block template repeated over and over in a certain fashion,
for ESO's telescope, such as the VST, a new OB is to be provided by the investigator using ESO's P2PP tool. Normally all Obs of a given semester need to be submitted to ESO by a certain deadline.
In our case, given changes in orbit due to corrective thrusts, we have agreed with ESO to deliver Obs on a monthly basis.     
\subsection{The internal structure of GBOT}
\label{sec:setup:structure}
\begin{figure}
   \begin{center}
   \includegraphics[height=8cm]{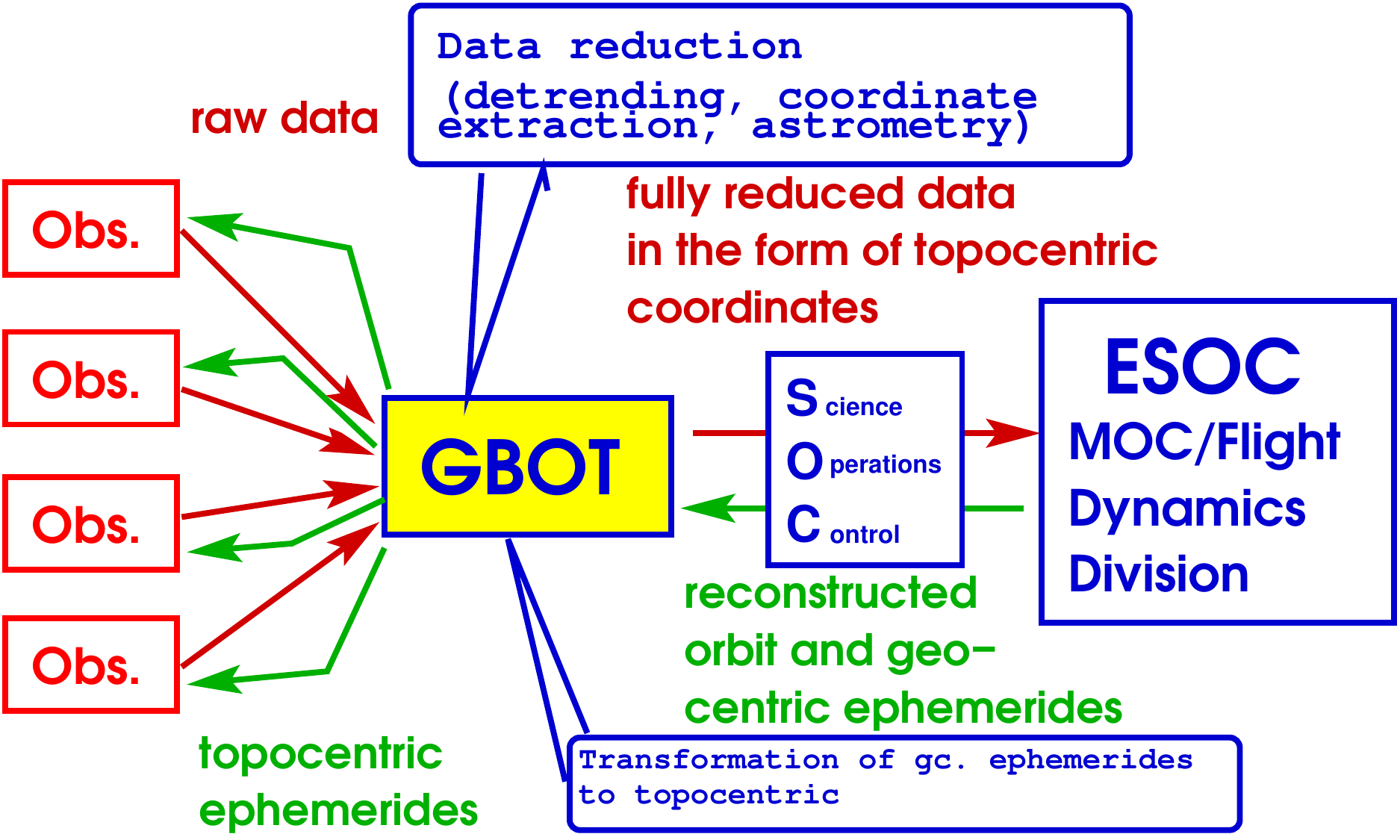}
   \end{center}
   \caption[GBOT Organigram]{
   \label{fig:organigram}
The Organigram of GBOT and its relations to the Gaia project and the observatories.
}
\end{figure}
The core of GBOT is located in two main locations, the {\it GBOT Data Processing Centre} at the Observatoire de Paris, run by members of the SYRTE group, and the {\it GBOT Office}, 
at the Astronomisches Recheninstitut in Heidelberg, Germany. In the GBOT Data Processing Centre the data processing and software development is being done. It is also the location of 
GBOT's main database. Both the data analysis and transformation of orbit files to topocentric ephemeris files are done here. The GBOT Office coordinates the effort, deals with strategic 
and operational issues, and represents the group to the outside, e.g. to the Gaia community, the Data Processing and Analysis Consortium (DPAC), of which GBOT is a part, the observatories, ESOC, ESAC, etc.
A somewhat less powerful mirror intended to keep a redundant backup and to continue operations in the event the main database in Paris is inoperable, is installed in Heidelberg. This infrastructure is
powerful enough to continue day to day operations and keep up the ephemeris server, but it is not equipped for the re-reduction process off all previous data after the 1st Gaia astrometry release.
However since the GBOT-data re-reduction is not a time critical process, this should not be a significant problem. The two mirrors are kept apart geographically on purpose, in order to prevent
a catastrophic event, such as a fire, destroying both mirrors.

The pipeline and database system including the ephemeris server are described in more detail in Bouquillon et al. (2014)\cite{Bouquillon_SPIE2014}. Therefore here we will give only a brief description. 
The pipeline has the task to cope with widely different data, import and harmonise this, e.g. by adding generic GBOT keywords to the header, extract the sources on the frame, perform an astrometric reduction using a reference catalogue (which can be chosen from a 
list),
measure the position of the moving object, and determine its celestial coordinates, and create a detailed set of diagnostic output. Various methods of centroiding are available both for the 
background stars as well as the moving target. Parameters like the order of the polynomials used to do the astrometric fits can be defined a priori. 
The results are then stored in a SAADA based 
database. The database can also be accessed by authorised persons from outside of Paris, giving access to the results to all members of the GBOT group. The various routines of the GBOT pipeline
are being run through the database, and the results, diagnostics as well as the incoming data are permanently stored in the database.   
\subsection{GBOT and the outside world: data deliveries and communications}
\label{sec:setup:outside}
Since GBOT is not a stand alone project, but a part of the Gaia mission, the official procedures GBOT has to follow are embedded in the context of DPAC procedures. 
The main consumer of GBOT's results is the Flight Dynamics group at the MOC (Mission Operations Control) located at ESOC in Darmstadt, 
Germany. However GBOT does not directly deliver the results, which are stored in a file 
(called an OPTO file) which has a file format mutually agreed on, to ESOC, but to a common interface, called SOC (Science Operations Centre), located at ESAC in Villafranca near Madrid, Spain.
This entity deals with all communications and deliveries between the various parts of the mission, such as GBOT, and ensures that these function in a standardised way. In a large space mission,
such as Gaia, a rigidly designed set of procedures are paramount - any sloppiness can easily lead to misunderstandings, and other problems. SOC then distributes the GBOT deliveries to MOC.
GBOT delivers apart from the OPTO file, which contains the astrometric results and the observing dates, an OPTT file, containing the telescope coordinates and other relevant data.
The orbit delivery from MOC to GBOT follows the same routine. 

Communications with the observatories are more informal, and depend on the procedures of the observatories. As described before, GBOT supplies the necessary observing instructions and ephemerides
to ESO through their Phase 2 system, observed data are being downloaded from the ESO archive. The other two main contributors use the GBOT Finding chart and ephemeris server in order to obtain
these information and insert it into their queue. The data observed by the LT are automatically downloaded from their archive, and LCOGT's frames from the IPAC repository hosted at Caltech.

Apart from the operational data and information transfer described above there is also need of verbal communication within the group, which is dispersed over several countries in and outside of Europe. 
This is accomplished by monthly teleconferences between GBOT and the observatories, and biannual or yearly meetings. This way the functioning of this group is secured, and all relevant issues are
taken care of.
\section{OPERATIONS and GBOT'S REASSESSMENT}
\label{sec:operations}
\begin{figure}
   \begin{center}
   \includegraphics[height=6cm]{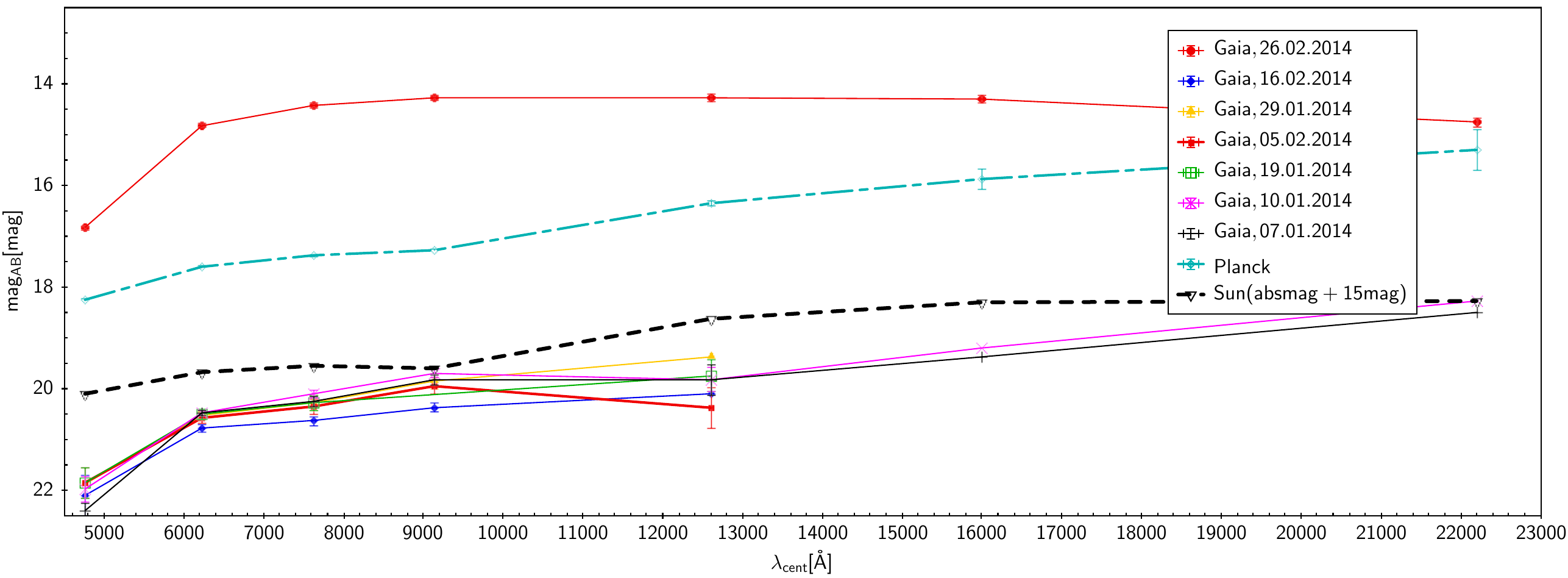}
   \end{center}
   \caption[GROND photometry of Gaia]{
   \label{fig:grond}
GROND multipassband optical and NIR photometry of Gaia in relation to the central wavelengths of the passbands, from blue to red: $g'$,$r'$,$i'$,$z'$,$J,H,K$. All magnitudes are in the AB system,
i.e. this plot gives a good representation of the SED of the light reflected off Gaia. For comparison photometry of ESA's Planck mission (diamonds, and dashed-dotted line) 
and the Sun (downward pointing open triangles and dashed line)
incremented by +15 mag are shown. The uppermost sequence shows Gaia (red symbols and line in the digital version of this paper)
during one of the Solar Aspect Angle = 0 diagnostic manoeuvres, taking place on Feb. 26, 2014, during which the space craft became significantly brighter ($\simeq6$ mags) than normally. 
Except for the solar values, all points have their photometric errors indicated, the $H$ and $K$ values for Gaia except during its bright phase are upper limits. 
}
\end{figure}

\begin{figure}
   \begin{center}
   \includegraphics[height=5cm]{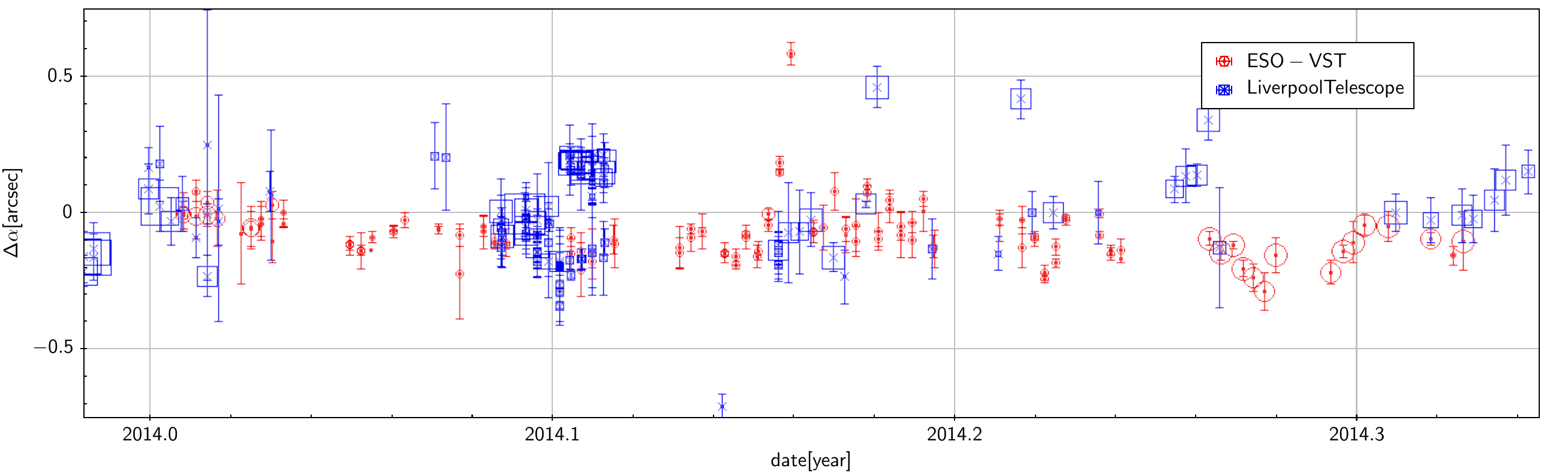}
   \includegraphics[height=5cm]{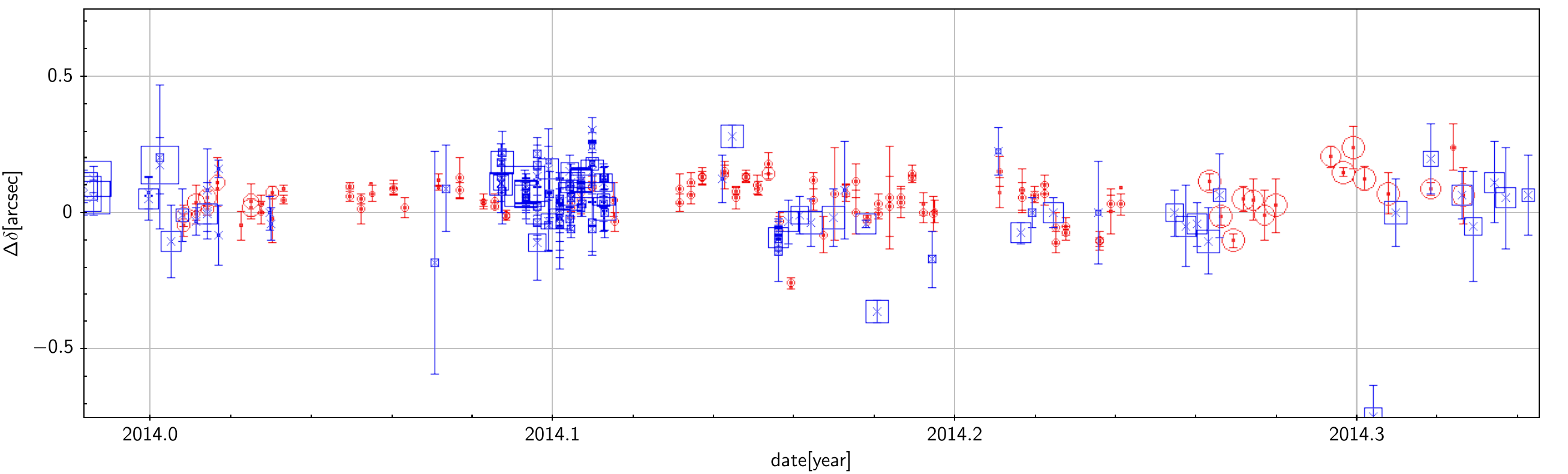}
   \includegraphics[height=5cm]{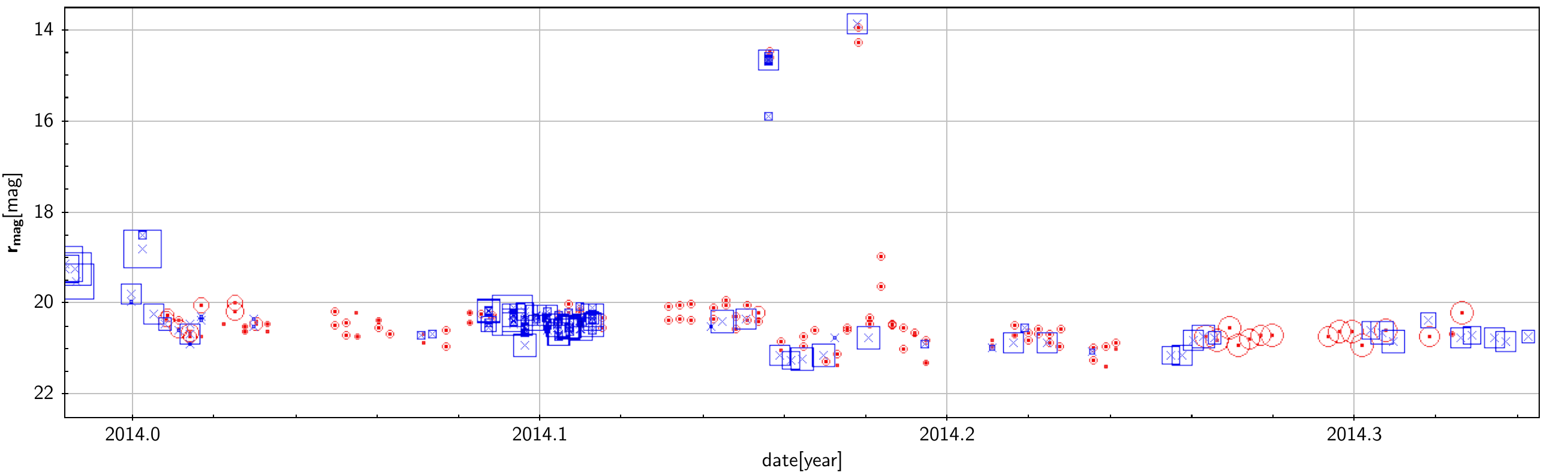}
   \end{center}
   \caption[GBOT Gaia measurements]{
   \label{fig:gaiadat}
The GBOT measurements since launch. The upper panel shows the $O-C$ residua for the right ascension, the middle panel for declination, and the lower panel the development
of the magnitude (mostly $r_{\rm SDSS}$). The red circular symbols (grey in the printed version) denote data taken with the VST, the blue crosses and squares (dark grey in the printed version) LT-data.
The errorbars show the scatter of a given sequence, the large symbols (open circles or squares) centred on the  small symbols (filled circles or crosses) are scaled according to the number of target
measurements (this may or may not be equal to the number of frames, since the target, i.e. Gaia is not always detected on every frame, or a measurement could be rejected for another reason), larger 
symbols meaning more detections. The $O-C$ values were derived using the reconstructed orbit which was  current at the time of observations (instead of calculating everything with the 
most recent orbit). This is presumably the main cause of the short term trends in the plotted $O-C$ values, lasting about one week, visible in the two upper panels.
}
\end{figure}

\begin{figure}
   \begin{center}
   \includegraphics[height=4.0cm]{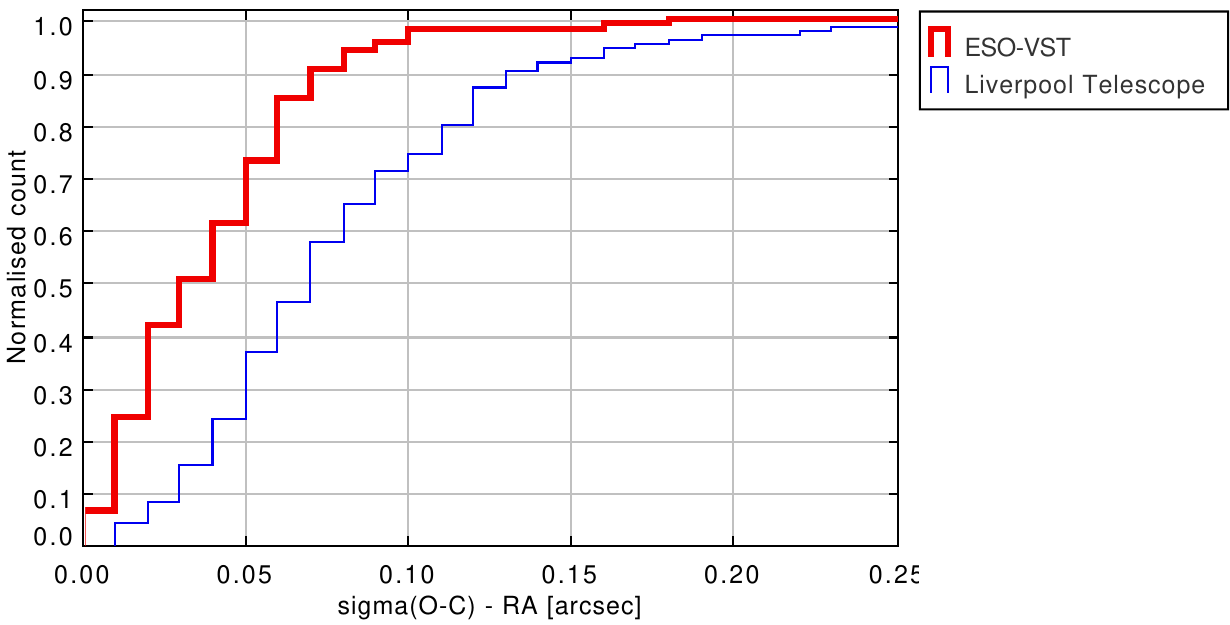}
   \includegraphics[height=4.0cm]{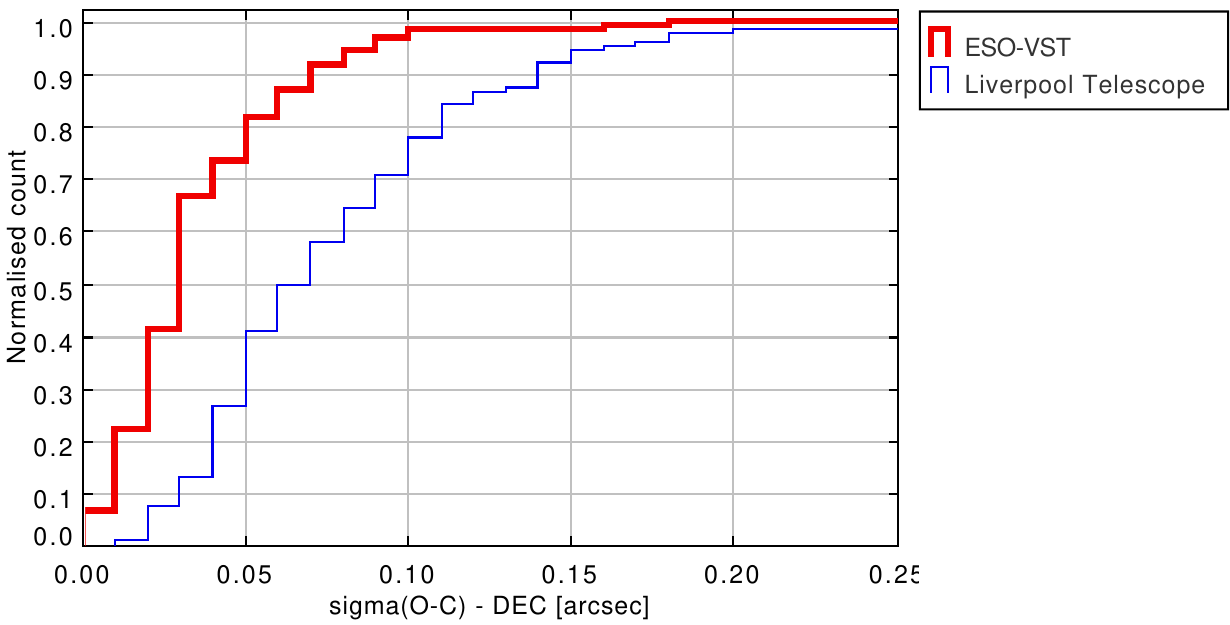}
   \end{center}
   \caption[GBOT Gaia astrometric scatter histograms]{
   \label{fig:gaiahist}
Cumulative normalised histograms of the scatter $\sigma(O-C)$ indicating the achievable precision. The thicker red curve (grey in the printed version) shows the distribution for the ESO-VST, and the
thinner blue curve (dark grey in the printed version) denotes the Liverpool telescope precision. In order to achieve the commitment precision of 20 mas, 
the cutoff $\sigma(O-C)$ for a standard sequence of 10 exposures is 63 mas, which can be reached in about 80-85\% of instances with the VST, and 50-60\% of the smaller Liverpool telescope. Therefore
it is prudent to increase the number of exposures per sequence on the latter to $\simeq15$, meaning a cutoff of 78 mas, which can be reached on 65-70\% of the observations. 
}
\end{figure}

Gaia was launched on December 19, 2013, and one day later slewed to its operational solar aspect angle of 45\degr. Soon it became clear, that the brightness assumptions, mainly based on experience
with other similar space probes, especially WMAP, which was from its outer shape a smaller version of Gaia, just with an operational solar aspect angle of 22.5\degr\, of about $R=$18 mag, was not
realistic. The final magnitude was now estimated to be 21.2 mag when in its final position in the L2-vicinity, leading to the urgent need to reevaluate the GBOT project for feasibility and reassess
the procedures and approaches. This much lower magnitude did not only call for more light gathering power, but also for an optimisation of our methods, since 21 mag is similar to the sky background for
most very good sites. This still ongoing process formed the main activities of the GBOT team during the opening months of 2014. Now, with several months having elapsed since the launch, we have a somewhat
better grasp of the magnitude range of Gaia; it was found to be between 20 and 21.2, depending on the distance, the Earth aspect angle\footnote{Since Gaia oscillates around the L2 point, and can deviate
from it by up to 15\degr\ tangentially, the Earth aspect angle and Solar aspect angle do in general not coincide}, and other factors. It was known before that space craft can show rapid and long term
variability of unknown origin, e.g. the unusual faintness followed by unusual brightness of Planck during the OR3 (see Sect. \ref{sec:observations}).
 
This unexpected situation has forced us to delve deeply into the theoretical foundations of astrometric measurements in order to determine whether we can still fulfil our promises despite the
faintness of Gaia. While this reassessment is not complete at present, some results have flowed into Sect. \ref{sec:theory}. This applies especially to the CRLB discussion, as elaborated by the study of 
Mend\'ez et al (2013)\cite{Mendez1}, which provides hard lower limits of what can be achieved under what conditions, i.e. sky brightness, FWHM, magnitude, see Fig. \ref{fig:CRLB}.
Since there are reasons to believe that Gaia's brightness could change in the long term (months or years) due to material degradation caused by the harsh radiative environment in the L2 region,
 we requested some photometric observations to be taken with the GROND
$b'r'i'z'JHK$ simultaneous imager mounted on the 2.2 m MPIA telescope on La Silla, see Fig. \ref{fig:grond}. These were kindly granted to us by GROND PI Jochen Greiner and reduced by Thomas Kr\"uhler,
thus far no unexpected long term trends in brightness have been found. Therefore we assume that Gaia will stay as faint as it is now.
 
At the same time, we are optimising the source extraction routines of our pipeline (see Bouquillon et al.,2014\cite{Bouquillon_SPIE2014}). Since later December/early January GBOT is gathering data 
on a regular basis mostly using the Liverpool telescope and ESO's VST. Judging from the results of this, we see that the VST regularly delivers data with an r.m.s. scatter 
consistent with our specifications,
assuming an observing sequence of 10 images. For the Liverpool telescope the number of exposures should be increased to 15, in which case the r.m.s. is within specifications in about 2/3 of the time.
While it is too early to come to a final conclusion on the situation, it is clear that GBOT can operate with some compromises, i.e. the number of nights delivering a precision beyond the
specifications will be higher, the Full Moon gap will be wider, more observations will fail in less than optimal conditions, and background stars will more often have a negative impact on the results. 
But, overall we will reach our aims. 

\section{OUTLOOK}
\label{sec:outlook}
The Gaia satellite itself is at current still in its commissioning phase, and will start nominal operation shortly, see Prusti (2014)\cite{Prusti_SPIE2014}. GBOT is already operating in nominal mode, alongside with the ongoing reassessment, and the data generated by GBOT will be used to enhance the orbit reconstruction.
From our intermediate results we can infer that we are operating near the limits of what can be done with moderate aperture telescopes. Should there ever be a need for another GBOT-like campaign,
we would give the advice that it is paramount to pay more attention to the brightness assumption, which is an issue far more complicated than it looks at first. While this paper is not intended to be
a cook book for satellite tracking campaigns, we do provide a number of hints and information, that hopefully makes the task of someone establishing a similar campaign in the future easier.   
\acknowledgments     

The authors would like to thank all people involved in helping the establishment of the GBOT project.
Specifically we would like to acknowledge the help in terms of test observing time, or help with various issues of the following people:
Ren\'e A. Mend\'ez, Todd Boroson, Tim Lister, Tim Brown, Jon Marchant, Monika Petr, Luciano Nicastro, Paolo Tanga, Richard Boyle, William Thuillot, Lukasz Wyrzykowski,
Jochen Greiner, Thomas Kr\"uhler, Laurent Eyer, Uli Bastian, Timo Prusti, Anthony Brown, Stefan Jordan, Werner Zeilinger, Francois Colas, 
Frank Budnik, Trevor Morley, Gabriele Bellei, Alejandro Lopez Lozano, and many more who have contributed to GBOT in one or another way.   
This work made use of data from ESO-programs 092.B-0165 and 093.B-0236, and the LCOGT network, and the Liverpool telescope.

\bibliography{report}   
\bibliographystyle{spiebib}   

\end{document}